\documentclass[showpacs,aps,prd,reprint,superscriptaddress]{revtex4-1}
\usepackage[colorlinks=true, pdfstartview=FitV, linkcolor=red,citecolor=blue, urlcolor=blue,
bookmarks=true, bookmarksnumbered=true, breaklinks]{hyperref}
\usepackage[dvipdfmx]{graphicx}

\usepackage{amsmath,amssymb,bm,ascmac,color,longtable,mathrsfs}

\usepackage{slashed}

\allowdisplaybreaks[1]

\usepackage{color}

\usepackage[normalem]{ulem}

\def\simge{\mathrel{
   \rlap{\raise 0.511ex \hbox{$>$}}{\lower 0.511ex \hbox{$\sim$}}}}
\def\simle{\mathrel{
   \rlap{\raise 0.511ex \hbox{$<$}}{\lower 0.511ex \hbox{$\sim$}}}}
\def\bigs{\mathrel{
   \rlap{\raise 0.531ex \hbox{$>$}}{\lower 0.531ex \hbox{$<$}}}}

\allowdisplaybreaks[3]
\allowdisplaybreaks

\begin{document}

\title{Kondo cloud of single heavy quark in cold and dense matter}

\author{Shigehiro~Yasui}
\email[]{yasuis@th.phys.titech.ac.jp}
\affiliation{Department of Physics, Tokyo Institute of Technology, Tokyo 152-8551, Japan}

\begin{abstract}
The Kondo effect is a universal phenomena observed in a variety of fermion systems containing a heavy impurity particle whose interaction is governed by the non-Abelian interaction.
At extremely high density, I study the Kondo effect by color exchange in quark matter containing a single heavy (charm or bottom) quark as an impurity particle.
To obtain the ground state with the Kondo effect, I introduce the condensate mixing the light quark and the heavy quark (Kondo cloud) in the mean-field approximation.
I estimate the energy gain by formation of the Kondo cloud, and present that the Kondo cloud exhibits the resonant structure.
I also evaluate the scattering cross section for the light quark and the heavy quark, and discuss its effect to the finite size quark matter.
\end{abstract}


\maketitle

\section{Introduction}
Heavy (charm or bottom) quarks, whose dynamics is governed by Quantum Chromodynamics (QCD), are useful tools to investigate the properties of nuclear and quark matter.
It is considered that the dynamics of the heavy quarks are hardly affected by light quarks, and the heavy-quark symmetry for heavy flavor and spin makes the dynamics simple; therefore they enable us to carry out a systematic analysis of the structures and reaction processes~\cite{Neubert:1993mb,Manohar:2000dt}. 
However, the situation may not be so simple when heavy quarks exist in medium at low temperature and high density.
In condensed matter physics, it is known that the medium properties can be changed significantly once the system is contaminated with impurities.
For instance, the Kondo effect occurs when a small number of impurity particles are injected into the system, and it changes the thermodynamic and transport properties~
\cite{Hewson,Yosida,Yamada}.
In this paper, I study the Kondo effect in quark matter at low temperature and high density, which is called the QCD Kondo effect~\cite{Yasui:2013xr,Hattori:2015hka,Yasui:2016svc}.
The research on the QCD Kondo effect has had positive spin-offs so far: the Kondo effect in strong magnetic field~\cite{Ozaki:2015sya} and in the color superconductivity~\cite{Kanazawa:2016ihl}, the Fermi/non-Fermi liquid properties in multi-channel Kondo effect~\cite{Kimura:2016zyv}, and non-trivial topological properties of heavy quark spin in ground state~~\cite{Yasui:2017izi}.
Also, it has been argued that one can analyze the information about high density matter, given the fact that heavy quarks can be produced in interior cores of neutron stars~\cite{Yasui:2016svc} or in heavy ion collisions at low energy in accelerator facilities such as GSI-FAIR~\cite{Friman:2011zz} and J-PARC.

In the 1960s, J.~Kondo established the theoretical foundations for the effect in order to explain why the electric resistance of a metal containing a small amount of magnetic (spin) impurities behaved with a logarithmical enhancement at low temperatures.
His research indicated that the spin-exchange (non-Abelian) interaction between the conducting electron and the spin impurities led to the situation where the coupling constant got stronger and the Landau pole appeared~\cite{Kondo:1964}.
The emergence of the pole was closely related to not only the non-Abelian interaction but also the Fermi surface (degenerate states) and the loop effect (virtual creation of particle-hole pairs).
So far, many theoretical methods to study the Kondo effect have been developed~\cite{Hewson,Yosida,Yamada}, and nowadays it is thought that the Kondo effect provides a direct insight into strong coupling mechanisms in condensed matter systems~\cite{Goldhaber-Gordon:1998,Cronenwett:1998,Wiel:2000,Jeong:2001,Park:2002,PhysRevLett.111.135301,Hur:2015}.

The Kondo effect can appear in the nuclear/quark matter containing heavy impurity particles (i.e. heavy hadrons/quarks); their energy scales are not comparable to that of the Kondo effect which involves the system of electrons.
Several types of non-Abelian interactions are used for the strong interaction: the spin exchange with SU(2) symmetry and the isospin exchange with SU(2) symmetry in nuclear matter~\cite{Yasui:2013xr,Yasui:2016ngy,Yasui:2016hlz}~\footnote{See Ref.~\cite{Sugawara-Tanabe:1979} for early application of the Kondo effect to deformed nuclei.}; color exchange with SU(3) symmetry in quark matter~\cite{Yasui:2013xr,Hattori:2015hka,Ozaki:2015sya,Yasui:2016svc,Kanazawa:2016ihl,Kimura:2016zyv}.
The Fermi surface and the loop effect naturally appear in nuclear/quark matter at low temperatures; these conditions seem to be favorable for the QCD Kondo effect to occur.

The enhancement of the interaction by the Kondo effect will cause changes in thermodynamic properties and transport properties in nuclear/quark matter.
Thus, the Kondo effect provides us with a tool to study (i) heavy-hadron--nucleon (heavy-quark--light-quark) interaction, (ii) modification of impurity properties by medium and (iii) change of nuclear/quark matter by impurity effect~(cf.~Ref.~\cite{Hosaka:2016ypm}).

So far, the QCD Kondo effect has been studied within the framework of perturbation, but the approach becomes invalid at low energy scales (the Kondo scale)~\cite{Yasui:2013xr,Hattori:2015hka,Ozaki:2015sya}.
Likewise the Landau pole appears in the scattering amplitude on Fermi surface at low temperatures, as the origin of Fermi instability which is known in superconductivity.
Below the Kondo scale, all the physical quantities have their poles, and so meaningful results cannot be obtained~(see Refs.~\cite{Hewson,Yosida,Yamada}).
Thus, the non-perturbative approach should be used to investigate their low-energy behaviors in terms of the QCD Kondo effect.

Recent studies dealt with this issue by using a non-perturbative approach and the mean-field approximation~\cite{Yasui:2016svc,Kanazawa:2016ihl}~\footnote{See Refs.~\cite{Yasui:2016ngy,Yasui:2016hlz} for application to charm/bottom nuclei}.
The mean-field approach is simple, but it gives us a general understanding of the Kondo effect, as it does in condensed matter physics~\cite{Takano:1966,Yoshimori:1970,Lacroix:1979,ReadNewns1983,Eto:2001,Yanagisawa:2015conf,Yanagisawa:2015}~\footnote{See also Ref.~\cite{Hewson} and references therein}.
The use of the mean-field approach can be justified, because it  gives the results consistent with the exact solutions in ground state in large $N$ for $\mathrm{SU}(N)$ symmetry in non-Abelian interaction term~\cite{ReadNewns1983}.
In Refs.~\cite{Yasui:2016svc,Kanazawa:2016ihl}, an infinitely extended matter state of heavy quarks was considered under the assumption that the number of the heavy quarks was infinite.
But those studies did not cover the situation where a single heavy quark could exist in quark matter which may be realized in heavy ion collisions at low energy.
In the present paper, I consider the system where there is a single heavy quark in quark matter, and use the mean-field approach to study its ground state.
I call an attention that the big difference between the previous studies and the present one is that in the latter, the existence of the heavy quark breaks the translational symmetry of the system, because the single heavy quark behaves like a point defect in quark matter.

In the mean-field approach, I define the Kondo cloud by the condensate of a light quark with a heavy quark, i.e. the light-hole--heavy -quark condensate and the light-quark--heavy-antiquark condensate in the color singlet channel~\cite{Yasui:2016svc}. 
The formation of the condensate makes the system stable. 
The previous study~\cite{Yasui:2016svc}, involving color current interaction between the two quarks, showed that the Kondo effect occurs on quark matter at low temperatures and high densities under the assumption that heavy quarks were uniformly distributed in space.
In the present paper, considering a single heavy quark in quark matter, I demonstrate that the Kondo cloud emerges as the resonant state.

The paper consists of six sections.
In section~\ref{sec:color_interaction}, I introduce the color current interaction for a light quark and a heavy quark.
Section~\ref{sec:perturbation} gives a brief description of the perturbation approach to a scattering of a light quark and a heavy quark, and shows that the approach becomes invalid at low energy scales.
In section~\ref{sec:Kondo_cloud}, I use the mean-field approach to investigate the Kondo cloud for a single heavy quark.
Section~\ref{sec:discussion} presents the numerical results of the cross section of the scattering of the light quark and the heavy quark in the Kondo cloud, and discuss the effects on transport property of quark matter under the influence of the Kondo cloud.
The final section is devoted to the conclusion.

\section{Color current interaction} \label{sec:color_interaction}

The interaction between the light quark ($\psi$) and the heavy quark ($\Psi$) is supplied by the gluon-exchange in QCD.
To simplify the discussion, I use the contact interaction with zero range instead of the gluon-exchange interaction with finite range~\cite{Klimt:1989pm,Vogl:1989ea,Klevansky:1992qe,Hatsuda:1994pi}.
The essence of the discussion does not change for the QCD Kondo effect, as long as color (non-Abelian) exchange is maintained in the interaction~\cite{Hattori:2015hka}.
For the simple setting, 
I regard the light quark mass zero, and regard the heavy quark mass infinity large.
As for the heavy quark, following the heavy-quark symmetry~\cite{Neubert:1993mb,Manohar:2000dt}, I consider the $v$-frame in which the heavy quark is at rest, and separate the heavy quark momentum as $p^{\mu}=m_{Q}v^{\mu}+k^{\mu}$ for the heavy quark mass $m_{Q}$, where $m_{Q}v^{\mu}$ is the on-mass-shell part ($v^{\mu}$ the four-velocity with $v^{\mu}v_{\mu}=1$) and the off-mass-shell part $k^{\mu}$ whose scale is much smaller than $m_{Q}$.
Correspondingly, I define the heavy quark effective field $\Psi_{v}=\frac{1+v\hspace{-0.4em}/}{2}e^{im_{Q}v\cdot x} \Psi$ by factorizing out the on-mass-shell part and leaving only the off-mass-shell part in the original field $\Psi$~\cite{Neubert:1993mb,Manohar:2000dt}.
Notice $\bar{\Psi}_{v}\Psi_{v}=\Psi_{v}^{\dag}\Psi_{v}$.

Based on the above setup,
I give the Lagrangian
\begin{eqnarray}
{\cal L}
&=&
\bar{\psi} (i \partial\hspace{-0.55em}/ + \mu\gamma^{0})\psi
+\bar{\Psi}_{v} v\!\cdot\!i\partial \Psi_{v}
\nonumber \\
&&
-G_{c} \sum_{a=1}^{N_{c}^{2}-1} \bar{\psi} \gamma^{\mu} T^{a} \psi \, \bar{\Psi}_{v} \gamma_{\mu} T^{a} \Psi_{v},
\label{eq:L}
\end{eqnarray}
with $\psi=(\psi_{1},\dots,\psi_{N_{f}})^{t}$ for the light flavor number $N_{f}$.
I will set $N_{f}=2$ in numerical calculation.
In the right hand side, the first and second terms represent the kinetic terms for light and heavy quarks, respectively, with $\mu$ the chemical potential for light quarks.
The third term is the interaction term with the coupling constant $G_{c}>0$.
$T^{a}=\lambda^{a}/2$ ($a=1,\dots,N_{c}^{2}-1$; $N_{c}=3$) are generators of color SU($N_{c}$) group with $\lambda^{a}$ the Gell-Mann matrices.
This color exchange interaction mimics the one-gluon exchange interaction.
It was shown that the one-gluon exchange interaction between a light quark and a heavy quark is screened by the Debye mass in finite-density medium, and that it leads to the short-range interaction~\cite{Hattori:2015hka}.
In the followings, I set $v^{\mu}=(1,\bm{0})$ as a static frame.

\section{Perturbative treatment of scattering} \label{sec:perturbation}

Based on Eq.~(\ref{eq:L}), let us investigate how the interaction strength can be changed by medium effect in quark matter.
When a heavy quark exists as an impurity particle, it can be dressed by virtual pairs of light quark and hole near the Fermi surface.
Due to those pairs, the interaction strength $G_{c}$ in Eq.~(\ref{eq:L}) can be modified from the value in vacuum.
I use the renormalization group method to estimate the value of the $G_{c}$ in medium.
Such analysis is known as the so called ``poor man's scaling" as an early application of renormalization group to the Kondo effect~\cite{Anderson1970}.
This method was also applied to the QCD Kondo effect with one-gluon exchange interaction~\cite{Hattori:2015hka}.
The following derivation essentially obeys the description in Ref.~\cite{Hattori:2015hka}. 

I consider the scattering process of a light quark $q$ and a heavy quark $Q$,
 $q_{l}(p) + Q_{j} \rightarrow q_{k}(p') + Q_{i}$,
where $p$ and $p'$ are in-coming and out-going four-momenta, and the subscripts $i,j,k,l=1, \dots, N_{c}$ stand for the color indices.
I consider that the interaction strength $G_{c}$ depends on the energy scale below or above the Fermi surface, and hence it is denoted by $G_{c}(\Lambda)$ for given energy scale $\Lambda$, which is measured from the Fermi surface.
The renormalization group equation leads to the relationship between $G_{c}(\Lambda-\mathrm{d}\Lambda)$ for lower scale $\Lambda-\mathrm{d}\Lambda$ and $G_{c}(\Lambda)$ for higher scale $\Lambda$, where $\mathrm{d}\Lambda>0$ is an infinitely small quantity.
The difference between $G_{c}(\Lambda-\mathrm{d}\Lambda)$ and $G_{c}(\Lambda)$ is given by the loop diagram with momentum integration in the small shell region between $\Lambda-\mathrm{d}\Lambda$ and $\Lambda$.

\begin{figure}[tb]
\begin{center}
\includegraphics[scale=0.35]{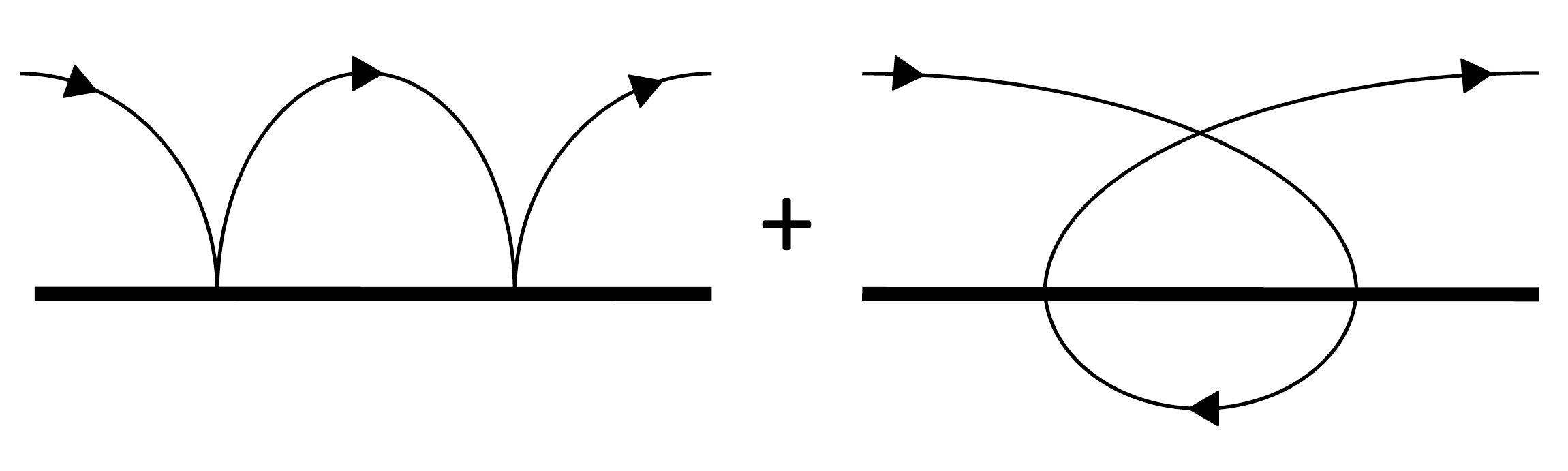}
\caption{The diagrams for scattering amplitude with one loop. Left: particle-propagating diagram. Right:hole-propagating diagram. The thin (thick) line indicates the propagator of a light (heavy) quark.}
\label{fig:Kondo}\vspace{-5mm}
\end{center}
\end{figure}

\begin{figure}[tb]
\begin{center}
\includegraphics[scale=0.4]{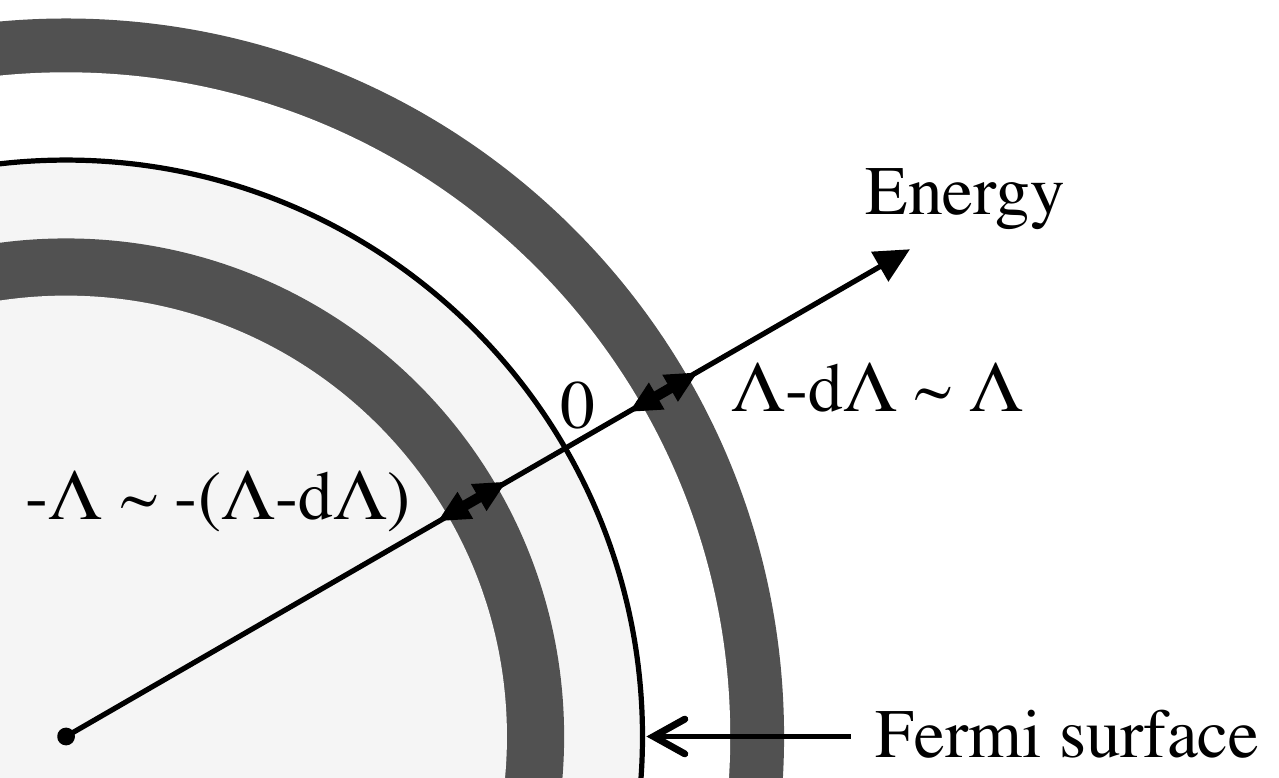}
\caption{The momentum region integrated in ${\cal M}^{(2)}$ (thick gray areas) in momentum space. The Fermi sphere is also shown (thin gray area).}
\label{fig:momentum}\vspace{-5mm}
\end{center}
\end{figure}

When I set $v^{\mu}=(1,\bm{0})$ in Eq.~(\ref{eq:L}),
I obtain the scattering amplitude at tree level,
\begin{eqnarray}
 {\cal M}^{(1)}_{\Lambda} = -iG_{c}(\Lambda) \gamma^{0} T_{kl,ij},
\end{eqnarray}
at certain energy scale $\Lambda$, where I define
 $T_{kl,ij} = \sum_{a=1}^{N_{c}} (T^{a})_{kl} (T^{a})_{ij}$
following the notation in Ref.~\cite{Hattori:2015hka}.
I can similarly introduce the scattering amplitude
\begin{eqnarray}
 {\cal M}^{(1)}_{\Lambda-\mathrm{d}\Lambda} = -iG_{c}(\Lambda-\mathrm{d}\Lambda) \gamma^{0} T_{kl,ij},
\end{eqnarray}
at lower energy scale $\Lambda-\mathrm{d}\Lambda$.
Now, as shown in Fig.~\ref{fig:Kondo}, I consider the scattering amplitude with one-loop where the momentum integration is supplied between $\Lambda-\mathrm{d}\Lambda$ and $\Lambda$. This is given by a sum of particle-propagating diagram and hole-propagating diagram, as
\begin{eqnarray}
 {\cal M}^{(2)}_{\Lambda-\mathrm{d},\Lambda}
&=&
 \bigl(-iG_{c}(\Lambda)\bigr)^{2} \int \frac{\mathrm{d}^{4}q}{(2\pi)^{4}} \gamma^{0} iS_{F}(q) \gamma^{0} 
\nonumber \\
&&\hspace{5em} \times
 \frac{i}{p_{0}-(q_{0}+\mu)+i\varepsilon}
 {\cal T}^{p}_{lk,ij}
\nonumber \\
&&
 +
 \bigl(-iG_{c}(\Lambda)\bigr)^{2} \int \frac{\mathrm{d}^{4}q}{(2\pi)^{4}} \gamma^{0} iS_{F}(q) \gamma^{0} 
\nonumber \\
&&\hspace{5em} \times
 \frac{i}{-p_{0}+(q_{0}+\mu)+i\varepsilon}
 {\cal T}^{h}_{lk,ij}.
\end{eqnarray}
Notice that the region of momentum integrals is limited to $[-\Lambda,-(\Lambda-\mathrm{d}\Lambda)]$ and $[\Lambda-\mathrm{d}\Lambda,\Lambda]$ below and above the Fermi surface, as shown in Fig.~\ref{fig:momentum}.
I define the propagator for a light quark with four-momentum $q^{\mu}=(q_{0},\bm{q})$ in matter
\begin{eqnarray}
 iS_{F}(q)
= q\hspace{-0.5em/}
\Bigl( \frac{i}{q^{2}+i\varepsilon} -2\pi \delta(q^{2}) \theta(q_{0}) \theta(\mu-|\bm{q}|) \Bigr),
\end{eqnarray}
with a chemical potential $\mu$ and an infinitely small and positive number $\varepsilon$.
The color matrices ${\cal T}^{p}_{lk,ij}$ for the particle-propagating diagram and ${\cal T}^{h}_{lk,ij}$ for the hole-propagating diagram
are defined by
\begin{eqnarray}
{\cal T}^{p}_{kl,ij}
= \frac{1}{2} \left( 1-\frac{1}{N_{c}^{2}} \right) \delta_{kl} \delta_{ij} - \frac{1}{N_{c}} T_{kl,ij},
\label{eq:Tp}
\end{eqnarray}
and 
\begin{eqnarray}
{\cal T}^{h}_{kl,ij} 
= \frac{1}{2} \left( 1-\frac{1}{N_{c}^{2}} \right) \delta_{kl} \delta_{ij} - \left( \frac{1}{N_{c}} -\frac{N_{c}}{2} \right)T_{kl,ij}.
\label{eq:Th}
\end{eqnarray}
Notice the second term in each of Eqs.~(\ref{eq:Tp}) and (\ref{eq:Th}) represents the non-Abelian property in the interaction.

Considering the finite density part, I obtain the renormalization group equation,
 ${\cal M}^{(1)}_{\Lambda-\mathrm{d}\Lambda} = {\cal M}^{(1)}_{\Lambda} + {\cal M}^{(2)}_{\Lambda-\mathrm{d}\Lambda,\Lambda}$,
which can be given by
\begin{eqnarray}
 \Lambda \frac{\mathrm{d}}{\mathrm{d}\Lambda} G_{c}(\Lambda) = -\frac{N_{c}}{8\pi^{2}} \mu^{2} G_{c}(\Lambda)^{2},
 \label{eq:renormalization}
\end{eqnarray}
for $G_{c}(\Lambda)$.
I suppose the value for $G_{c}(\Lambda_{0})$ at the initial energy scale $\Lambda_{0}$ which is much far from Fermi surface is almost same with the value in vacuum; $G_{c}(\Lambda_{0})=G_{c}$.
This assumption would be justified because the medium effect should be turned off away from the Fermi surface. 
Then, from Eq.~(\ref{eq:renormalization}), I obtain the $\Lambda$-dependence of $G_{c}(\Lambda)$
\begin{eqnarray}
 G_{c}(\Lambda) = \cfrac{G_{c}}{1+ \cfrac{N_{c}}{8\pi^{2}} \mu^{2} G_{c} \ln \cfrac{\Lambda}{\Lambda_{0}}},
\end{eqnarray}
at the energy scale $\Lambda$.
However, the above solution cannot reach the low energy limit $\Lambda=0$.
This is because the value of $G_{c}(\Lambda)$ becomes divergent below the low energy scale $\Lambda_{K} (\ll \Lambda_{0})$, where $\Lambda_{K}$ is defined by
\begin{eqnarray}
 \Lambda_{K} = \Lambda_{0} \exp\Biggl( - \cfrac{8\pi^{2}}{N_{c}\mu^{2}G_{c}} \Biggr).
\end{eqnarray}
This is the Landau pole in the infrared energy region~\footnote{Notice that the enhancement behavior of $G_{c}(\Lambda)$ originates from minus sign in r.h.s. in Eq.~(\ref{eq:renormalization}), where the origin of this minus sign can be found in the non-cancellation of $T_{kl,ij}$ in ${\cal T}^{p}_{kl,ij}-{\cal T}^{h}_{kl,ij}$ as the non-Abelian property.}.
It indicates that the perturbation theory cannot be applied for smaller energy scale $\Lambda < \Lambda_{K}$.
As a result, I cannot obtain the effective coupling strength which includes fully the virtual particle-hole pairs at the Fermi surface.
Hence, it is impossible for us to calculate the scattering cross section of a light quark and a heavy quark in ground state.
To solve this difficulty, I need to rely on non-perturbative treatment for the Kondo effect,
 as discussed in the next section.

\section{Kondo cloud for heavy quark} \label{sec:Kondo_cloud}
\subsection{Color singlet channel}

As the non-perturbative treatment, I consider the mean-field approximation.
For this purpose, first of all, I introduce a condition that the number density of the heavy quark is concentrated at the position $\bm{x}=\bm{0}$,
\begin{eqnarray}
 \bar{\Psi}_{v}\Psi_{v}\!=\!\delta^{(3)}(\bm{x}).
\end{eqnarray}
This condition can be accounted naturally by modifying the Lagrangian (\ref{eq:L}) as
\begin{eqnarray}
 {\cal L}_{\lambda}
=
{\cal L} - \lambda \bigl(\bar{\Psi}_{v}\Psi_{v}\!-\!\delta^{(3)}(\bm{x}) \bigr),
\end{eqnarray}
where the parameter $\lambda$ is introduced as the Lagrange multiplier.
For applying the mean-field approximation, I adopt the Fierz transformation~\footnote{I use
$
\sum_{a=1}^{N_{c}} (T^{a})_{ij} (T^{a})_{kl} =8\delta_{il}\delta_{kj}-({8}/{N_{c}})\delta_{ij}\delta_{kl},
$
for color matrices.}, and
rewrite the interaction term as a sum of 
\begin{eqnarray}
 2(\bar{\psi}_{\ell}\gamma^{\mu}\Psi_{v})(\bar{\Psi}_{v}\gamma_{\mu}\psi_{\ell}),
\end{eqnarray}
and
\begin{eqnarray}
 -\frac{2}{N_{c}} (\bar{\psi}_{\ell}\gamma^{\mu}\psi_{\ell})(\bar{\Psi}_{v}\gamma_{\mu}\Psi_{v}),
\end{eqnarray}
for each light flavor $\ell=1,\dots,N_{f}$ including the sign by interchanging two fermion fields.
I neglect however the latter term because it is irrelevant to the color exchange.
For the former term, I perform the mean-field approximation as
\begin{eqnarray}
\bar{\psi}_{\ell\alpha}\Psi_{v\delta}\bar{\Psi}_{v\gamma}\psi_{\ell\beta}
&\rightarrow&
\langle \bar{\psi}_{\ell\alpha} \Psi_{v\delta} \rangle \bar{\Psi}_{v\gamma} \psi_{\ell\beta}
+ \langle \bar{\Psi}_{v\gamma} \psi_{\ell\beta} \rangle \bar{\psi}_{\ell\alpha} \Psi_{v\delta}
\nonumber \\
&&
- \langle \bar{\psi}_{\ell\alpha} \Psi_{\delta} \rangle \langle \bar{\Psi}_{v\gamma} \psi_{\ell\beta} \rangle,
\end{eqnarray}
for the mean-field $\langle \bar{\psi}_{\ell\alpha} \Psi_{\delta} \rangle$, whose value is obtained by the self-consistent calculation or by minimizing the total energy~\footnote{This is regarded as the loop expansion in large $N_{c}$. See also the discussion of the large $N_{c}$ limit in Sec.~\ref{sec:competition}.}.
I define the gap function 
\begin{eqnarray}
\hat{\Delta}^{\bm{1}}_{\ell\delta\alpha} &=& \frac{G_{c}}{2} \langle \bar{\psi}_{\ell\alpha} \Psi_{\delta} \rangle,
\end{eqnarray}
in which the Dirac indices $\delta$, $\alpha$ can be factorized,
so that $\hat{\Delta}^{\bm{1}}_{\ell\delta\alpha}=\Delta^{\bm{1}}_{\ell} ( \frac{1+\gamma_{0}}{2} ( 1-\hat{\bm{k}}\!\cdot\!\bm{\gamma} ) )_{\delta \alpha}$ with $\hat{\bm{k}}=\bm{k}/|\bm{k}|$ for three-momentum of the light quark $\bm{k}$.
I introduce the complex scalar quantity $\Delta^{\bm{1}}_{\ell}$ to measure the size of the gap function. 
I set $\Delta^{\bm{1}}_{\ell}=\Delta_{\bm{1}}$ for all $\ell$ by assuming the light flavor symmetry.
In the mean-field approximation, the Lagrangian is simplified to the bilinear form of $(\psi,\Psi_{v})$,
and hence the energy eigenvalues of the system can be calculated straightforwardly.
As a result, I obtain the total energy (thermodynamic potential) $\Omega_{\bm{1}}(\lambda,\Delta_{\bm{1}})$ as a function of $\lambda$ and $\Delta_{\bm{1}}$, whose values are determined by the stationary condition for $\Omega_{\bm{1}}(\lambda,\Delta_{\bm{1}})$:
\begin{eqnarray}
 \frac{\partial}{\partial \lambda} \Omega_{\bm{1}}(\lambda,\Delta_{\bm{1}})=\frac{\partial}{\partial \Delta_{\bm{1}}^{\ast}} \Omega_{\bm{1}}(\lambda,\Delta_{\bm{1}})=0.
\end{eqnarray}

The thermodynamic potential of the heavy quark is
\begin{eqnarray}
\Omega_{\bm{1}}(T,\lambda,\Delta_{\bm{1}})
&=& -\frac{1}{\beta} \int_{-\infty}^{+\infty} \!\! \ln \! \left( 1\!+\!e^{-\beta \omega} \right) 2N_{c} \rho_{\bm{1}}(\omega) \mathrm{d}\omega
\nonumber \\
&&+ \frac{8N_{f}|\Delta_{\bm{1}}|^{2}}{G_{c}} V
- \lambda,
\label{eq:potential_1_T}
\end{eqnarray}
with $\beta=1/T$, the inverse temperature,
where $\rho_{\bm{1}}(\omega)$ is the density-of-state of the heavy quark
\begin{eqnarray}
\hspace*{-1em}
\rho_{\bm{1}}(\omega)=-\frac{1}{\pi} \mathrm{Im} \, \frac{\partial}{\partial \omega} \ln \! \left( \omega_{+} \!-\! \lambda \!-\! \sum_{\bm{k}} \frac{2N_{f}|\Delta_{\bm{1}}|^{2}}{\omega_{+}\!+\!\mu\!-\!|\bm{k}|} \right),
\label{eq:density_1}
\end{eqnarray}
as a function of the energy $\omega$ measured from the Fermi surface with $\omega_{+}=\omega+i\eta$ for $\eta$ an infinitely small and positive number.
The coefficient $2N_{c}$ in the integral in Eq.~(\ref{eq:potential_1_T}) represents the number of degrees of freedom of heavy quark, spin and color,
and $V$ is the volume of the system.
The derivation of Eq.~(\ref{eq:density_1}) is presented in Appendix~\ref{sec:dos}.
Notice that the sum for $\bm{k}$ in Eq.~(\ref{eq:density_1}) is taken over all momenta with different sizes and directions, and it represents the coherent sum of the light quarks which couple to the heavy quark.
This reflects the non-conservation of the light quark three-momenta in scattering by the impurity particle.

I calculate the thermodynamic potential (\ref{eq:potential_1_T}) in the following.
I introduce the cutoff parameter $\Lambda$ in the $\omega$-integral, and restrict the integral range to $[-\Lambda,\Lambda]$ around the Fermi surface.
This procedure is necessary for normalizing the integral due to the zero-range interaction in the present Lagrangian.
Considering zero temperature as low-temperature limit ($\beta\rightarrow\infty$), I find that Eq.~(\ref{eq:potential_1_T}) can be represented by a simple form
\begin{eqnarray}
\Omega^{0}_{\bm{1}}(\lambda,\delta_{\bm{1}})
&=&
\frac{2N_{c}}{\pi} \left( -\delta_{\bm{1}} + \lambda \arctan \! \frac{\delta_{\bm{1}}}{\lambda} + \frac{\delta_{\bm{1}}}{2} \ln \! \frac{\lambda^{2}+\delta_{\bm{1}}^{2}}{\Lambda^{2}} \right)
\nonumber \\
&&+\frac{8\pi}{\mu^{2} G_{c}} \delta_{\bm{1}}
-\lambda,
\label{eq:potential_1_T0}
\end{eqnarray}
with the approximation
\begin{eqnarray}
\sum_{\bm{k}} \frac{2N_{f}|\Delta_{\bm{1}}|^{2}}{\omega_{+}+\mu-|\bm{k}|}
&\simeq& -\frac{iN_{f}}{\pi} \mu^{2} |\Delta_{\bm{1}}|^{2} V
\nonumber \\
&& \equiv -i\delta_{\bm{1}},
\end{eqnarray}
in Eq.~(\ref{eq:density_1})
 by neglecting the real part and leaving the imaginary part only.
Then, Eq.~(\ref{eq:density_1}) becomes
\begin{eqnarray}
 \rho_{\bm{1}}(\omega) = \frac{1}{\pi} \frac{\delta_{\bm{1}}}{(\omega-\lambda)^{2}+\delta_{\bm{1}}^{2}},
\end{eqnarray}
which represents the resonant state by Lorentzian with energy position $\omega=\lambda$ and width $\Gamma_{\bm{1}}=2\delta_{\bm{1}}$.
This resonance is called the Kondo cloud. It can be regarded as a composite state of the light quark and the heavy quark because of the mixing effect by $\delta_{\bm{1}}$ (or $\Delta_{\bm{1}}$).

From the stationary condition for $\Omega_{\bm{1}}^{0}(\lambda,\Delta_{\bm{1}})$, I obtain $\delta_{\bm{1}}$ and $\lambda$ as
\begin{eqnarray}
 \delta_{\bm{1}} = \Lambda \sin\! \left( \frac{\pi}{2N_{c}} \right) \exp\! \left( -\frac{4\pi^{2}}{N_{c}\,\mu^{2}G_{c}} \right),
 \label{eq:delta1}
\end{eqnarray}
and $\lambda=\delta_{\bm{1}}/ \tan\! \left( {\pi}/{2N_{c}} \right)$.
By substituting them to Eq.~(\ref{eq:potential_1_T0}), I obtain the thermodynamic potential
\begin{eqnarray}
\Omega^{0}_{\bm{1}} = -\Lambda \frac{2N_{c}}{\pi} \sin\! \left( \frac{\pi}{2N_{c}} \right) \exp\! \left( -\frac{4\pi^{2}}{N_{c}\,\mu^{2}G_{c}} \right),
\label{eq:pot1}
\end{eqnarray}
in the ground state.
This is the energy gain by formation of the Kondo cloud.

\begin{figure}[tb]
\begin{center}
  \vspace*{-3em}
  \begin{minipage}[b]{1.0\linewidth}
    \centering
    \includegraphics[keepaspectratio, scale=0.28]{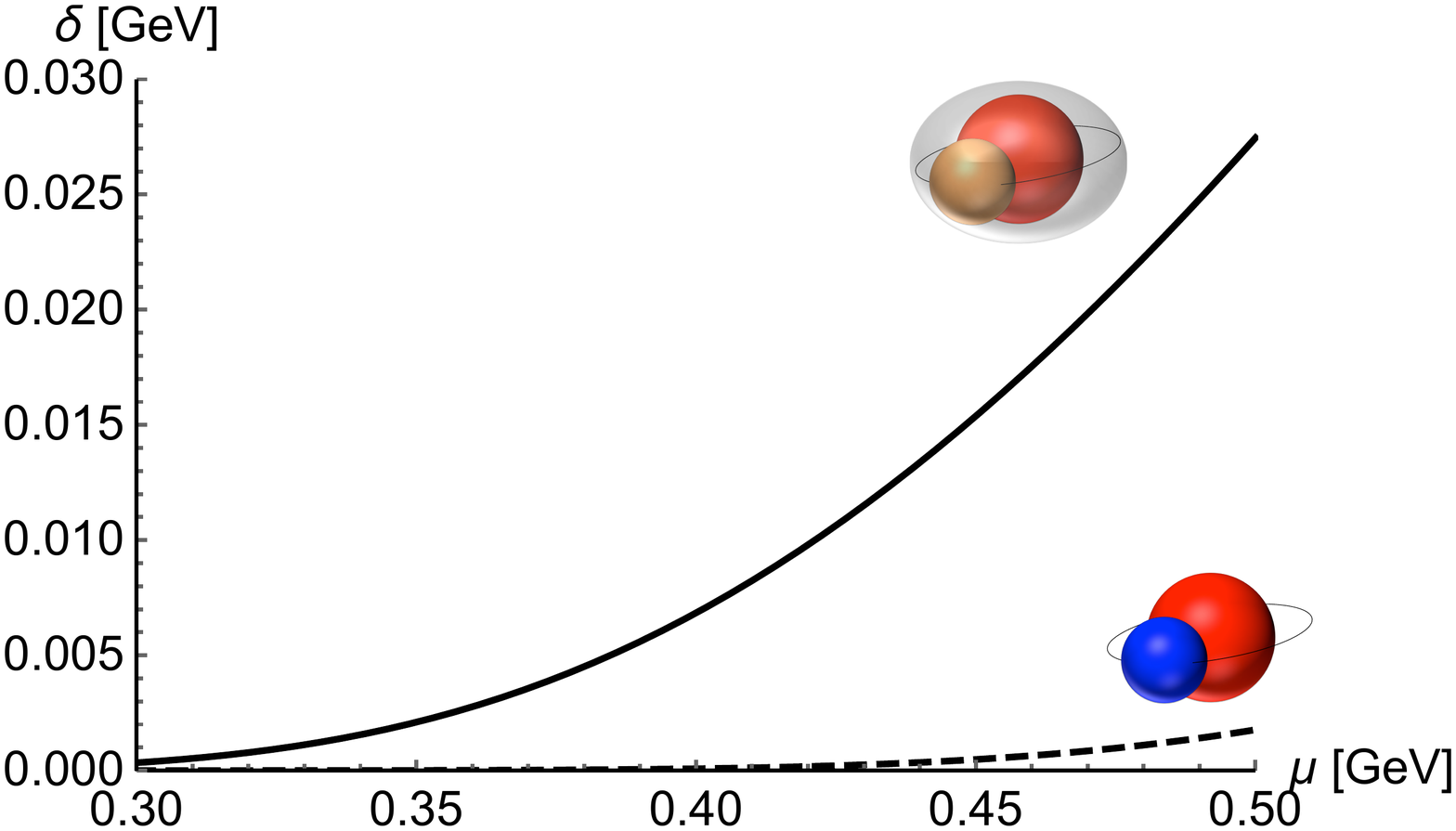}
  \end{minipage}\\
  \vspace*{-3em}
  \begin{minipage}[b]{1.0\linewidth}
    \centering
    \includegraphics[keepaspectratio, scale=0.28]{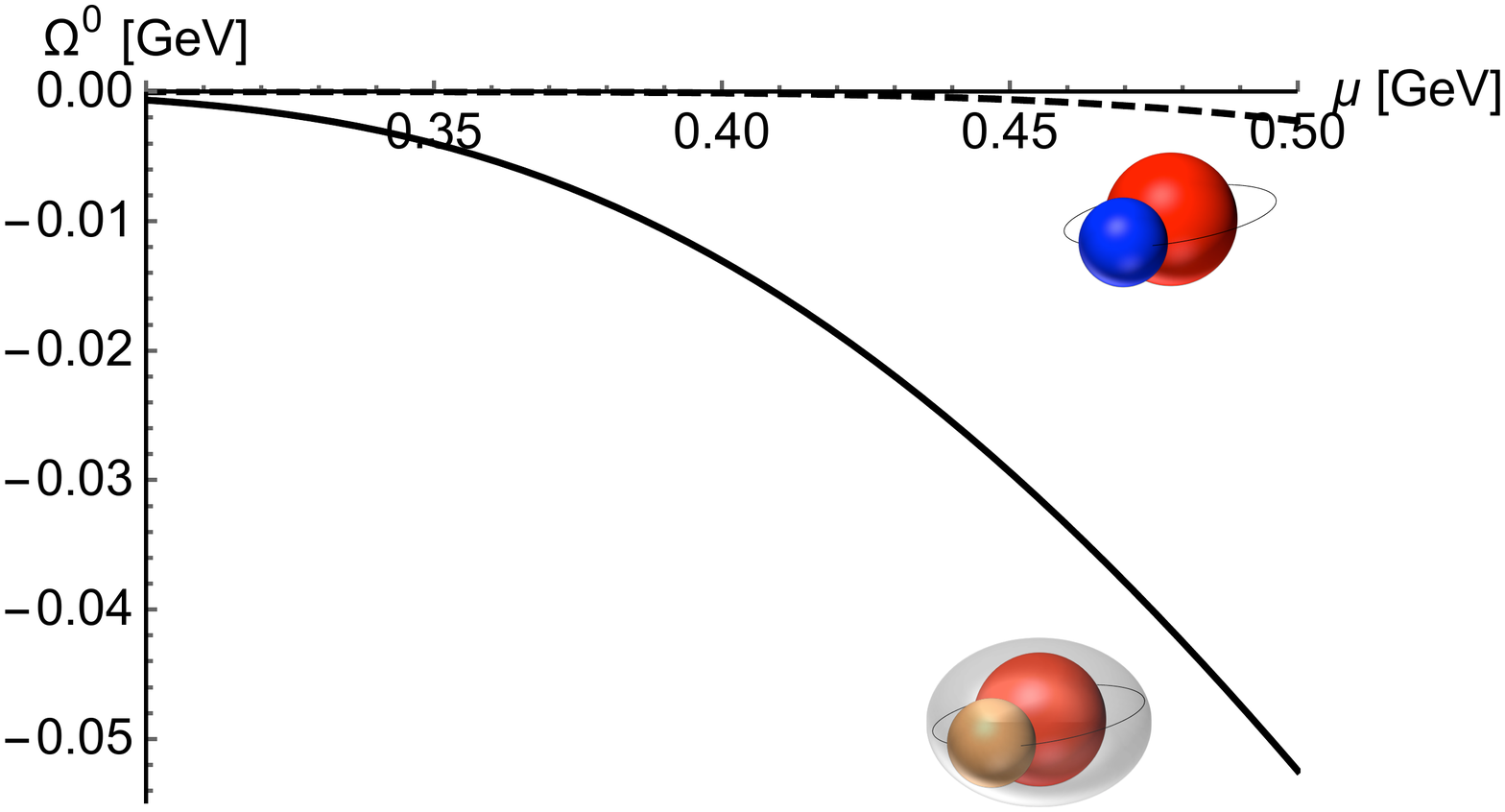}
  \end{minipage}
    \vspace*{-3em}
  \caption{Upper panel: the gaps $\delta_{\bm{1}}$ (solid curve) and $\delta_{\bar{\bm{3}}}$ (dashed curve) as functions of $\mu$. Lower panel: the thermodynamics potentials $\Omega_{\bm{1}}^{0}$ (solid curve) and $\Omega_{\bar{\bm{3}}}^{0}$ (dashed curve) as functions of $\mu$.}
  \label{fig:delta_pot}
  \end{center}\vspace{-5mm}
\end{figure}

For the numerical estimate, I use the coupling constant and cutoff parameter, $G_{c}=(9/2)2.0/\Lambda^{2}$ and $\Lambda=0.65$ GeV, in the color current interaction in Eq.~(\ref{eq:L}).
Those numbers are fixed to reproduce the chiral condensate and the pion decay constant in vacuum by the Nambu--Jona-Lasinio (NJL) interaction~\cite{Nambu:1961tp,Nambu:1961fr}, which is obtained by the Fierz transformation from the color current interaction~\cite{Klimt:1989pm,Vogl:1989ea,Klevansky:1992qe,Hatsuda:1994pi}.
The numerical results of $\delta_{\bm{1}}$ and $\Omega^{0}_{\bm{1}}$ are shown in Fig.~\ref{fig:delta_pot}.
For example, I obtain $\delta_{\bm{1}}= 0.027$ GeV ($\lambda=0.048$ GeV) and $\Omega^{0}_{\bm{1}}=-0.052$ GeV at $\mu=0.5$ GeV, where the binding energy of the color singlet Kondo cloud is regarded as 52 MeV.

\subsection{Color anti-triplet channel}
Instead of the color singlet channel, I may consider the color non-singlet channel.
I investigate the color anti-triplet Kondo cloud, i.e. the light-particle--heavy-quark condensate and the light-hole--heavy-antiquark condensate, as an attractive channel in the interaction in Eq.~(\ref{eq:L}).
Adopting the Fierz transformation\footnote{
To find the color anti-triplet interaction term, I use the identity
$\sum_{a=1}^{N_{c}} (T^{a})_{ij} (T^{a})_{kl}
= (1-1/N_{c}) \sum_{a:\rm{S}} (T^{a})_{ik}(T^{a})_{lj} 
- (1+1/N_{c}) \sum_{a:\rm{AS}} (T^{a})_{ik}(T^{a})_{lj}$,
where S (AS) stands for the (anti)symmetric term for interchanging two of the color indices $i,j,k,l=1,\dots,N_{c}$.},
I consider the color anti-triplet channel 
\begin{eqnarray}
 (\bar{\psi}_{\ell\alpha} \lambda^{2} \bar{\Psi}^{t}_{v\gamma}) (\Psi^{t}_{v \delta} \lambda^{2} \psi_{\ell\beta}),
\end{eqnarray}
for each light flavor $\ell=1,\dots,N_{f}$.
The other antisymmetric terms with $\lambda^{5}$ and $\lambda^{7}$ can be changed to the one with $\lambda^{2}$ by unitary transformation.
I apply the mean-field approximation as
\begin{eqnarray}
\bar{\psi}_{\ell\alpha} \lambda^{2} \bar{\Psi}^{t}_{v\gamma} \Psi^{t}_{v\delta} \lambda^{2} \psi_{\ell\beta}
&\rightarrow&
\langle \bar{\psi}_{\ell \alpha} \lambda^{2} \bar{\Psi}^{t}_{v\gamma} \rangle \Psi^{t}_{v\delta} \lambda^{2} \psi_{\ell \beta}
\nonumber \\
&& + \langle \Psi^{t}_{v\delta} \lambda^{2} \psi_{\ell \beta} \rangle \bar{\psi}_{\ell \alpha} \lambda^{2} \bar{\Psi}^{t}_{v\gamma}
\nonumber \\
&& - \langle \bar{\psi}_{\ell \alpha} \lambda^{2} \bar{\Psi}^{t}_{v\gamma} \rangle \langle \Psi^{t}_{v\delta} \lambda^{2} \psi_{\ell \beta} \rangle,
\end{eqnarray}
including the sign by interchanging two fermion fields.
I define the gap function
\begin{eqnarray}
\hat{\Delta}^{\bar{\bm{3}}}_{\ell\gamma\alpha} &=& \left( 1+\frac{1}{N_{c}} \right)
 \frac{G_{c}}{4} \langle \bar{\psi}_{\ell \alpha} \lambda^{2} \bar{\Psi}^{t}_{v\gamma} \rangle,
\end{eqnarray}
in which the Dirac indices $\delta$, $\alpha$ can be factorized,
so that $\hat{\Delta}^{\bar{\bm{3}}}_{\ell\gamma\alpha}=\Delta_{\ell}^{\bar{\bm{3}}} ( \frac{1+\gamma_{0}}{2} ( 1-\hat{\bm{k}}\!\cdot\!\bm{\gamma} ) )_{\gamma \alpha}$
with the complex scalar quantity $\Delta_{\ell}^{\bar{\bm{3}}}$.
I set $\Delta_{\ell}^{\bar{\bm{3}}}=\Delta_{\bar{\bm{3}}}$ for all $\ell$ by the light flavor symmetry.
I notice that the color symmetry is broken from SU($N_{c}$) to SU($N_{c}-1$).
This is analogous to the two-flavor diquark condensate in color superconductivity~\cite{Buballa:2003qv,Alford:2007xm,Fukushima:2010bq,Fukushima:2013rx}.

The thermodynamic potential of the heavy quark is
\begin{eqnarray}
&&\Omega_{\bar{\bm{3}}}(T,\lambda,\Delta_{\bar{\bm{3}}})
\nonumber \\
&=& -\frac{1}{\beta} \int_{-\infty}^{+\infty} \!\! \ln \! \left( 1\!+\!e^{-\beta \omega} \right) \!
\Bigl(4 \rho_{\bar{\bm{3}}}(\omega) + 2(N_{c}-2)\rho'_{\bar{\bm{3}}}(\omega) \Bigr) \mathrm{d}\omega
\nonumber \\
&&+ \frac{4N_{f}|\Delta_{\bar{\bm{3}}}|^{2}}{(1+1/N_{c})G_{c}} V
- \lambda,
\label{eq:potential_3_T}
\end{eqnarray}
where the density-of-state of the heavy quark is given by
\begin{eqnarray}
\hspace*{-2em}
\rho_{\bar{\bm{3}}}(\omega)
= -\frac{1}{\pi} \mathrm{Im} \, \frac{\partial}{\partial \omega} \ln \! \left( \omega_{+} \!-\! \lambda \!-\! \sum_{\bm{k}} \frac{2N_{f}|\Delta_{\bar{\bm{3}}}|^{2}}{\omega_{+}\!+\!\mu\!-\!|\bm{k}|} \right),
\label{eq:density_3}
\end{eqnarray}
and 
$
\rho_{\bar{\bm{3}}}'(\omega)
= 
-\frac{1}{\pi} \mathrm{Im} \, \frac{\partial}{\partial \omega} \ln \! \left( \omega_{+} - \lambda \right)
$.
The $\rho_{\bar{\bm{3}}}(\omega)$ is the heavy quark state which couples to the light quarks, and the $\rho_{\bar{\bm{3}}}'(\omega)$ is the one which does not couple.
In Eq.~(\ref{eq:potential_3_T}), there are the coefficients by spin and color, $4$ for $\rho_{\bar{\bm{3}}}(\omega)$ (e.g. red, green for color SU(2) symmetry) and $2(N_{c}-2)$ for $\rho_{\bar{\bm{3}}}'(\omega)$.

Similarly to the color singlet case, 
I restrict the $\omega$-integral range to $[-\Lambda,\Lambda]$ in Eq.~(\ref{eq:potential_3_T}).
Considering the zero temperature ($\beta \rightarrow \infty$), I obtain the simple form
\begin{eqnarray}
\Omega^{0}_{\bar{\bm{3}}}(\lambda,\delta_{\bar{\bm{3}}})
&=&
\frac{4}{\pi} \left( -\delta_{\bar{\bm{3}}} + \lambda \arctan \! \frac{\delta_{\bar{\bm{3}}}}{\lambda} + \frac{\delta_{\bar{\bm{3}}}}{2} \ln \! \frac{\lambda^{2}+\delta_{\bar{\bm{3}}}^{2}}{\Lambda^{2}} \right)
\nonumber \\
&&
+\frac{16\pi}{(1+1/N_{c})\mu^{2} G_{c}} \delta_{\bar{\bm{3}}}
-\lambda,
\label{eq:potential_3_T0}
\end{eqnarray}
for Eq.~(\ref{eq:potential_3_T})
with the approximation
\begin{eqnarray}
\sum_{\bm{k}} \frac{2N_{f}|\Delta_{\bar{\bm{3}}}|^{2}}{\omega_{+}+\mu-|\bm{k}|}
&\simeq& -\frac{iN_{f}}{\pi} \mu^{2} |\Delta_{\bar{\bm{3}}}|^{2} V
\nonumber \\
&\equiv& -i\delta_{\bar{\bm{3}}},
\end{eqnarray}
in Eq.~(\ref{eq:density_3}).
By the stationary condition for $\Omega^{0}_{\bar{\bm{3}}}$, I obtain $\delta_{\bar{\bm{3}}}$ and $\lambda$ as
\begin{eqnarray}
 \delta_{\bar{\bm{3}}} = \frac{\Lambda}{\sqrt{2}} \exp\! \left( -\frac{4\pi^{2}}{(1+1/N_{c})\,\mu^{2}G_{c}} \right),
 \label{eq:delta3}
\end{eqnarray}
and
$\lambda=\delta_{\bar{\bm{3}}}$.
By substituting them to Eq.~(\ref{eq:potential_3_T0}), I obtain the thermodynamic potential
\begin{eqnarray}
\Omega^{0}_{\bar{\bm{3}}}
 = -\Lambda \frac{2\sqrt{2}}{\pi} \exp\! \left( -\frac{4\pi^{2}}{(1+1/N_{c})\,\mu^{2}G_{c}} \right),
\label{eq:pot3}
\end{eqnarray}
 in the ground state.
Adopting the same parameter values used in the color singlet case, I obtain the numerical results of $\delta_{\bar{\bm{3}}}$ and $\Omega^{0}_{\bar{\bm{3}}}$ as shown in Fig.~\ref{fig:delta_pot}.
For $\mu=0.5$ GeV, for example, I obtain $\delta_{\bar{\bm{3}}}=0.0018$ GeV ($\lambda=0.0018$ GeV) and $\Omega^{0}_{\bar{\bm{3}}}=-0.0023$ GeV, where the binding energy of the color anti-triplet Kondo cloud is regarded as 2.3 MeV.

\subsection{Competition between color singlet and anti-triplet Kondo clouds} \label{sec:competition}
It is an interesting problem to ask which channel of the Kondo cloud is realized in the ground state, the color singlet or the color anti-triplet.
In Fig.~\ref{fig:delta_pot}, I find that the size of the color singlet gap $\delta_{\bm{1}}$ (or $|\Delta_{\bm{1}}|$) is larger than that of the color anti-triplet gap $\delta_{\bar{\bm{3}}}$ (or $|\Delta_{\bar{\bm{3}}}|$).
It indicates that the mixing strength of the light-hole--heavy-quark (or the light-particle--heavy-antiquark) in the color singlet channel is stronger than that of the light-quark--heavy-quark (or the light-hole--heavy-antiquark) in the color anti-triplet channel.
This is seen also in the stability of the Kondo cloud. 
In Fig.~\ref{fig:delta_pot}, the thermodynamic potential of the color singlet Kondo cloud is lower than that of the color anti-triplet Kondo cloud, and hence the former is realized in the ground state.

The above conclusion is consistent with what is expected in the large $N_{c}$ limit.
In this limit, generally, 
 the color singlet condensate survives, while the color non-singlet condensate vanishes~\cite{'tHooft:1973jz}.
In fact, by comparing Eq.~(\ref{eq:delta1}) to Eq.~(\ref{eq:delta3}), I find that the latter is suppressed by the small factor $\sim e^{-N_{c}}$ against the former 
 in the 't~Hooft limit, i.e. taking the large $N_{c}$ limit by keeping $g^{2}N_{c} \sim G_{c}N_{c}$ unchanged for the quark-gluon coupling constant $g$~\cite{'tHooft:1973jz}.
Hence, I confirm that the color singlet channel is more dominant than the color anti-triplet one.

\begin{figure}[tb]
\begin{center}
\includegraphics[scale=0.4]{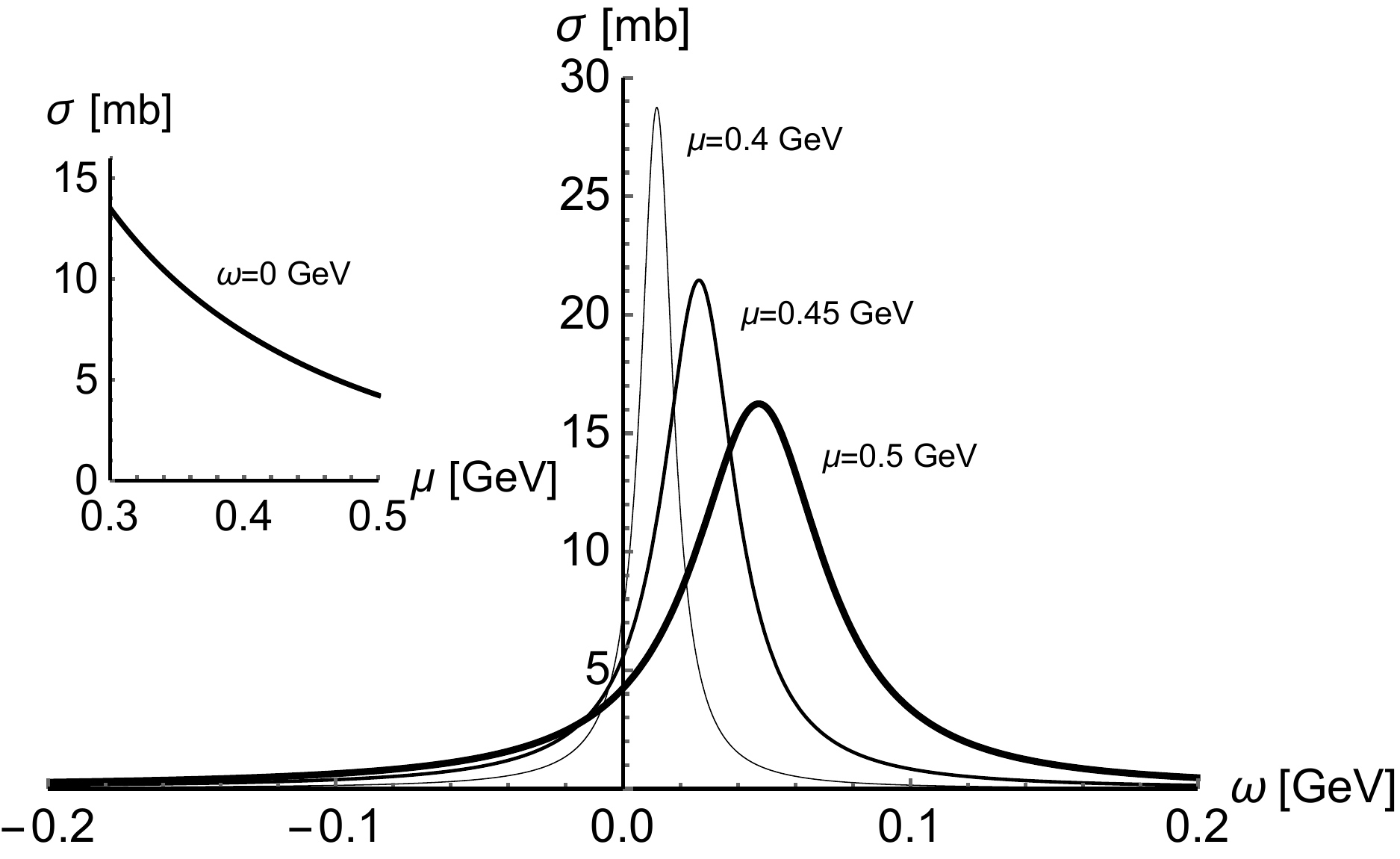}
\caption{The cross sections as functions of $\omega$ for $\mu=0.4$ GeV (thin line), $0.45$ GeV (normal line) and $0.5$ GeV (thick line). The behavior as a function of $\mu$ at $\omega=0$ GeV is also shown.}
\label{fig:crosssection}\vspace{-5mm}
\end{center}
\end{figure}

\section{Cross section} \label{sec:discussion}
The Kondo cloud affects the scattering of the light quark by the heavy quark.
For the color singlet Kondo cloud, I calculate the cross section and estimate the mean-free path in the quark matter.

The phase shift of the scattering 
 is given by
\begin{eqnarray}
 \Delta \delta(\omega) = \pi \int_{-\Lambda}^{\omega} \rho_{\bm{1}}(\omega') \mathrm{d}\omega'.
\end{eqnarray}
The scattering amplitude is
\begin{eqnarray}
 f(\omega) = \frac{1}{k} e^{i\Delta \delta(\omega)}\sin \Delta \delta(\omega),
\end{eqnarray}
with momentum $k=\mu+\omega$, and the cross section is
$
 \sigma(\omega) = 4\pi | f(\omega) |^{2}
$.
The numerical results are shown in Fig.~\ref{fig:crosssection}.
The resonance structure which represents the Kondo cloud can be seen clearly.
In low-temperature limit, the quarks with $\omega \simeq 0$ GeV on the Fermi surface dominantly contributes to the scattering process.
The cross section at $\omega=0$ GeV is presented in Fig.~\ref{fig:crosssection}.
The value of $\sigma(0)$ ranges from 13.6 mb to 4.2 mb for $\mu=0.3-0.5$ GeV.

It may be interesting to compare the obtained cross section with the cross section estimated in perturbative QCD at finite density.
I consider the one-gluon exchange with dynamical screening mass $m_{D} \simeq g\mu$ with a coupling constant $g$ for the quark-gluon vertex and quark chemical potential $\mu$.
Using the running coupling constant (e.g. one-loop order) at energy scale $\mu$,
I find that the cross section in one-gluon exchange $\sigma_{\mathrm{pQCD}}$ is much smaller than the values obtained by the Kondo cloud,
though a precise estimate is out of scope of the present work.

Transport coefficients are one of the most important quantities for study of the Kondo effect in experiments.
The quark matter at low temperature and high baryon number density may be produced in heavy ion collisions at low energy in GSI-FIAR and possibly in J-PARC.
There, the transport coefficients will be reflected in fluid dynamics such as elliptic flow $v_{2}$, nuclear suppression factor $R_{AA}$ and so on.
Here, I will leave a discussion about possible enhancement of the transport coefficients by the Kondo cloud.
Under the influence of the Kondo cloud, the cross section of the scattering of a heavy quark and a light quark $\sigma$ is much larger than the cross section by pQCD $\sigma_{\mathrm{pQCD}}$.
Hence, it is naively expected that the transport coefficients can be changed by the large factor $\sigma/\sigma_{\mathrm{pQCD}}  \gg 1$.
As an analogous situation, 
 the transport coefficients in quark-gluon plasma at high temperature are affected by the resonant states in medium~\cite{vanHees:2004gq,vanHees:2007me}.
The similar phenomena can occur for the Kondo cloud as a resonance state.

Based on the estimated cross section, I evaluate the mean-free-path
$
\tau \simeq (\sigma \, n_{q}/N_{f})^{-1}
$ with
 $n_{q}=\mu^{3}/\pi$ the number density of light quarks including the factor of spin, light flavor and color.
Notice that the number density per one flavor ($n_{q}/N_{f}$) contributes, 
 because the coherent sum is taken for the light flavor (cf.~Eq.~(\ref{eq:density_1})).
As a result, I obtain $\tau \simeq 1.3-0.91$ fm for $\mu=0.3-0.5$ GeV.
Those values are comparable with the mean-distance of a light quark, $\ell_{q} \simeq n_{q}^{-1/3}=0.96-0.58$ fm for $\mu=0.3-0.5$ GeV.
The mean-free-path provided by the Kondo cloud gives us an important hint 
to study the Kondo effect in heavy ion collisions in accelerator facilities.
The spatial size of the quark matter, that can be produced in relativistic heavy ion collisions, would be the same order as the initial nucleus size.
The obtained value of $\tau$ would be smaller than the size of the quark matter, hence the Kondo cloud can affect the properties 
 of the finite size quark matter in experiments.

\section{Conclusion}
I consider the Kondo effect in the quark matter containing a single heavy quark,
and investigate the Kondo cloud as the condensate of the light quark and the heavy quark in the ground state.
In the present analysis, I introduce the energy spectral function by considering the violation of the translational invariance of the system.
This formalism is different from the one used in Ref.~\cite{Yasui:2016svc}, where translational invariance was maintained due to the existence of matter state of heavy quarks.
I conclude that the color singlet Kondo cloud is realized, where the Kondo cloud is represented by the resonant state and the spectral function is given by the Lorentzian form.
As a study for application, I estimate the cross section for scattering of a light quark and a heavy quark by the Kondo cloud, and discuss the possible modification in transport property in quark matter.

To investigate the Kondo effect in quark matter is an interesting topics for future experimental studies.
The Kondo effect could be a good signal to search a high baryon number density.
To calculate the transport coefficients at quantitative level is an important problem, because they can become larger value due to the large cross section by the Kondo cloud.
The competition between superconductivity and Kondo effect is also an interesting subject~\cite{Kanazawa:2016ihl}~\footnote{See e.g. Ref.~\cite{RevModPhys.78.373} as a review paper in condensed matter physics.}.
The Kondo cloud may be closely related to the heavy hadron production in relativistic heavy ion collisions, where
either of a heavy quark and a heavy antiquark would be dressed in the color singlet Kondo cloud.
As one scenario, the bound state like $c\bar{c}$ would resolve, and it leads to suppression of the $J/\psi$ yield.
As another scenario, because the heavy (anti)quark accompanies the light hole (quark) in the Kondo cloud, it should affect the coalescence process for the hadronization of the heavy (anti)quark, leading to the modifications of heavy hadron yields.
They are left for future studies.

\section*{Acknowledgments}
The author thanks K.~Itakura and K.~Suzuki for fruitful discussions and thanks also M.~Oka for useful comments and T.~Miyamoto for careful reading of the manuscript.
The author thanks the Yukawa Institute for Theoretical Physics, Kyoto University, 
where this work was performed partially during the YITP-W-16-01 ``MIN16 - Meson in Nucleus 2016 -".
This work is supported by the Grant-in-Aid for Scientific Research (Grant No.~25247036, No.~15K17641 and No.~16K05366) from Japan Society for the Promotion of Science (JSPS).

\appendix

\section{Derivation of density of state in Eq.~(\ref{eq:density_1})}
\label{sec:dos}
I show the derivation of the density of state in Eq.~(\ref{eq:density_1}) along the prescription in Ref.~\cite{Hewson}.
For simple illustration, I assume the light flavor $N_{f}=1$, and extend it to multi-flavor in the end.
For the Hamiltonian of a single light quark with three-dimensional momentum $\bm{k}$,
\begin{eqnarray}
H^{0}_{\bm{k}} = 
\left(
\begin{array}{cc}
 -\mu & \bm{k}\!\cdot\!\bm{\sigma}   \\
 \bm{k}\!\cdot\!\bm{\sigma} & -\mu
\end{array}
\right),
\end{eqnarray}
the Hamiltonian in mean-field approximation is given by
\begin{eqnarray}
{\cal H}=
\left(
\begin{array}{cccc}
 H^{0}_{\bm{k}} \delta_{\bm{k},\bm{k}'} & \dots & \hat{\Delta}^{\dag}_{\bm{k}}  \\
 \vdots & \ddots  & \vdots  \\
 \hat{\Delta}_{\bm{k}} & \dots  &  \lambda
\end{array}
\right),
\end{eqnarray}
with $\hat{\Delta}_{\bm{k}} = ( -\Delta, -\Delta \hat{\bm{k}}\!\cdot\!\bm{\sigma} )$, where $\bm{k}$ ($\bm{k}\,'$) runs over all the momenta of the light quark as denoted by ellipses.
Notice that the different momenta are mixed by the off diagonal component, $\hat{\Delta}_{\bm{k}}$.
The non-conservation of the light quark momenta is a natural consequence of the existence of the heavy quark impurity located at the origin of spatial coordinate, because the existence of the point-like defect violates the translational invariance of the system.
I define the Green's function ${\cal G}(\omega)$ by
\begin{eqnarray}
 (\omega+i\eta-{\cal H}) {\cal G}(\omega) = 0,
\end{eqnarray}
for energy $\omega$ ($\eta$ an infinitely positive number).
I define the spectral function
\begin{eqnarray}
 \rho(\omega) = -\frac{1}{\pi} \mathrm{Im} \mathrm{Tr} \, {\cal G}(\omega),
\end{eqnarray}
where $\mathrm{Tr}$ is taken over all the light quark momenta.
Subtracting the free light quark part which is irrelevant to the heavy quark,
I obtain the density state of the heavy quark, including the mixing effect between the light quark and the heavy quark,
\begin{eqnarray}
\hspace*{-2em}
 \rho'(\omega)
  = - \frac{1}{\pi} \mathrm{Im} \frac{\partial}{\partial\omega}
\ln \left( \omega_{+} - \lambda-\sum_{\bm{k}}\cfrac{2|\Delta|^{2}}{\omega_{+}+\mu-|\bm{k}|} \right),
\end{eqnarray}
with $\omega_{+}=\omega+i\eta$.
The factor two for chirality is divided in $\rho(\omega)$.
It is straightforward to extend the above calculation to any light flavor $N_{f}$.
I finally obtain Eq.~(\ref{eq:density_1}). 

\bibliography{reference}

\end{document}